\documentstyle[a4,12pt]{article}

\newcommand{\sect}[1]{\setcounter{equation}{0}\section{#1}}
\renewcommand{\theequation}{\thesection.\arabic{equation}}

\newcommand{\NP}[1]{Nucl. Phys.\ {\bf #1}\ }
\newcommand{\PL}[1]{Phys. Lett.\ {\bf #1}\ }

\newcommand{\PR}[1]{Phys. Rev\ {\bf #1}\ }
\newcommand{\PRL}[1]{Phys. Rev. Lett.\ {\bf #1}\ }

\newcommand{\CQG}[1]{Class. Quantum Grav.\ {\bf #1}\ }
\newcommand{\PRep}[1]{Phys. Rept \ {\bf #1}\ }

\newcommand{\ba}{\begin{eqnarray}}
\newcommand{\ea}{\end{eqnarray}}
\newcommand{\be}{\begin{equation}}
\newcommand{\ee}{\end{equation}}
\newcommand{\lbl}{\label}

\begin{document}
\begin{titlepage}
\begin{flushright}
NBI-HE 96-28\\
June 1996\\
\end{flushright}
\vspace*{0.5cm}
\begin{center}
{\bf
\begin{Large}
Exact Solutions to Supergravity\\ 
\end{Large}
}
\vspace*{3cm}
         {\large A. K. Tollst\'{e}n\footnote{email tollsten@nbi.dk}}
         \\[.5cm]
         The Niels Bohr Institute, Blegdamsvej 17,\\
         DK-2100 Copenhagen \O, Denmark
\end{center}
\vspace*{2cm}
\begin{abstract}
We find a general $p-1$-brane solution to supergravity coupled to a $p+1$-form
field strength using the ``standard ansatz'' for the fields. In addition to the 
well-known elementary and solitonic $p-1$-brane solutions, which are the only
ones preserving half of the supersymmetry (for the cases where the 
supersymmetry
transformations are known), there are other possibilities, the  most 
interesting 
of which is an 
elementary type~I string solution with a non-dynamical string source and no
conserved charge.
\end{abstract}
\end{titlepage}
\sect{Introduction}
Many people now believe that all string theories are different perturbative
approximations of an, as yet, unknown theory. One consequence of this 
conjecture is that the strong coupling limit of a certain theory, with respect
to a certain coupling constant, is the weak coupling limit of another 
theory. Candidates for these dual pairs have been found for (practically)
all ten- and eleven-dimensional theories, and there are many interesting
examples in lower dimensions.
In such a strong-weak dual pair the fundamental states of one 
theory should 
occur as solitonic states of the dual one. 

There is much evididence that in ten dimensions the $SO(32)$ heterotic string 
is dual to the type~I string. 
Their low energy effective field theories are the same (after a field
redefinition which, among other things, changes sign of the dilaton, thus also 
exchanging the coupling constant $g=<e^\phi>$ with $1/g$) \cite{W95}, and 
some higher 
order terms have also been shown to agree after the same field redifinition
\cite{Ts95}.
There is a ``miraculous breakdown'' of type~I perturbation theory
whenever the toroidally compactified  heterotic string has an enhanced 
symmetry not present in the type~I string \cite{PW95}. And finally, the 
type~I string 
has the heterotic string as an exact classical solution \cite{Dab95,H195,PW95}.
However, so far  nobody has found the corresponding type~I state in the 
effective theory for the heterotic string. This is not so very surprising, 
since the standard classical $p-1$-brane solutions which have been examined
in this context, correspond to an infinite $p-1$-brane with constant energy 
per $p-1$-volume. For theories with only closed $p-1$-branes this is no
problem. The solution can be interpreted as a $p-1$-brane winding around
$p-1$ toroidally compactified coordinates, with very large compactification
radii (see \cite{Dab95}), but the type~I string theory also contains open 
strings, so such a state obviously cannot be stable here. The best 
one can do without radically changing one's ansatz is to look for a metastable
configuration around such an unstable type~I string explicitly put into 
the equations by hand.

In all, a careful study of exact solutions to low energy effective theories 
is certainly merited. In previous work on general $p-1$-brane solutions, 
one has always assumed supersymmetry (or rather, 
a generalization of the relations which tell us
that half the supersymmetries remain unbroken for the cases where the fermionic
theory and the supersymmetry transformations are known), see, for instance,
 \cite{DGHR90,DKL94}. In this paper we 
will solve the supergravity equations of motion without prejudice. Given the
``standard ansatz'' for the fields, the complete exact solution turns out to 
exist in closed form. Putting this solution back into the equations of motion,
however, gives us singularities, which we 
choose to put at the origin of the transverse coordinates. If we want the
equations of motion to be consistent also at the origin, we have to choose
the integration constants in a particular fashion. We can introduce a dynamical
$p-1$-brane source at the origin in which case we find the (supersymmetric)
elementary $p-1$-brane solution, or we can study the dual theory with no 
source, in which we find the (supersymmetric) solitonic $p-1$ brane.
These well-known cases appear to be the only supersymmetric ones. As far as
we know all the new solutions are configurations with a fixed source at the 
origin.

We will do the string case in great detail, keeping the metric general (all
possible rescalings with  $e^\phi$ included). We then find that one such 
fixed source solution exhibits the correct scaling behaviour to be the 
missing unstable type~I solution. In the end of the paper we also sketch the
case of generic dimensions, without trying to find any further possible 
interpretation.

\sect{The solution of ten-dimensional supergravity}
Our starting point is the bosonic part of the combined string supergravity
action in ten dimensions, here written in arbitrary metric 
\footnote{See Appendix for
notation and conventions}
\ba
\lbl{1}
\lefteqn{
S = \frac{1}{\kappa^2}\int d^{10}x 
\sqrt{-g}e^{4\left( a+\frac{1}{3}\right)\phi}
\left[ R-9a \left( a+\frac{2}{3}\right)\partial\phi^2+\frac{3}{2}
e^{-2\left( a+\frac{4}{3}\right)\phi}H^2\right]}
\nonumber\\
&& \hspace{-2mm}
-\frac{T_2}{2}\int d^2\xi \left[
\sqrt{-\gamma}\gamma^{ij}\partial_iX^M
\partial_j X^Ne^{\left( b+\frac{4}{3}\right)\phi}g_{MN}+2\varepsilon^{ij}
\partial_iX^M\partial_jX^NB_{MN}\right].
\ea
Here $a=-\frac{4}{3}$ gives us the usual (heterotic) string metric,
$a=-\frac{1}{3}$ the Einstein metric, and $a=\frac{2}{3}$ the type I string 
metric. Choosing $a=0$ gives us the five-brane metric, and if we also 
replace $H$ by its dual, $\tilde{H} = e^{2\left(a-\frac{2}{3}\right)\phi}*H$ 
we
obtain the action postulated to be the low energy effective action for 
five-brane theory \cite{DL291}.

For $b=a$ we have a heterotic string source, while
$b=a-2$ and, as it will turn out, no $B_{MN}$ term in the string part of 
the action will
give us a type~I source. Since we only consider solutions
with the Yang-Mills fields identically zero, we can omit the $\mbox{tr}{F^2}$
term in the action just like we do with all the fermionic terms. 

The variation of (\ref{1}) with respect to $g_{MN}$, $B_{MN}$, $\phi$,
$\gamma_{ij}$, and $X^M$, gives us the supergravity equations of motion, with
$\delta$-function sources at the location of the string, 
and the equations of motion for the
string degrees of freedom. For the supergravity fields 
we obtain Einstein's equation
\ba
\lbl{1g}
\lefteqn{
\sqrt{-g}e^{4\left( a+\frac{1}{3}\right)\phi}
\left\{ -R^{MN}-2\left( a+\frac{1}{3}\right)D^M\partial^N\phi\right.
+\left( a-\frac{1}{3}\right)\left( a+\frac{4}{3}\right)
\partial^M\phi\partial^N\phi }
\nonumber\\
&&\hspace{-.5cm}
-\frac{9}{2}e^{-2\left( a+\frac{4}{3}\right)\phi}H^MH^N
+\frac{g^{MN}}{2}\left[R+4\left( a+\frac{1}{3}\right)\Box\phi
+\left(7a\left( a+\frac{2}{3}\right)+\frac{16}{9}\right) \partial\phi^2\right.
\nonumber\\
&&\hspace{-.5cm}
+\left.\left.\frac{3}{2}e^{-2\left( a+\frac{4}{3}\right)\phi}H^2\right]\right\}
\nonumber\\ &&\hspace{-.5cm}
-\frac{\kappa^2 T_2}{2}\int d^2\xi \sqrt{-\gamma}\gamma_{ij}
\partial_iX^M
\partial_j X^Ne^{\left( b+\frac{4}{3}\right)\phi}\delta^{10}(X-x)=0,
\ea
the antisymmetric tensor equation
\ba
\lbl{1b}
-3\partial_M\left(\sqrt{-g}e^{2\left(a-\frac{2}{3}\right)\phi}H^{MNP}\right)\
-\kappa^2T_2\int d^2\xi\varepsilon^{ij}\partial_iX^M\partial_jX^N
\delta^{10}(X-x)=0,
\ea
and the dilaton equation
\ba
\lbl{1fi}
\lefteqn{\sqrt{-g}e^{4\left( a+\frac{1}{3}\right)\phi}}
\nonumber\\
&&\cdot  \left[4\left( a+\frac{1}{3}
\right)R+18a\left( a+\frac{2}{3}\right)\left(\Box\phi+2\left( a+
\frac{1}{3}\right)\partial\phi^2\right)+3\left( a-\frac{2}{3}\right)
H^2\right]
\nonumber\\
&&- \frac{\kappa^2T_2}{2}\int d^2\xi
\sqrt{-\gamma}\gamma^{ij}\partial_iX^M
\partial_j X^Ne^{\left( b+\frac{4}{3}\right)\phi}g_{MN}
\left( b+\frac{4}{3}\right)\delta^{10}(X-x)=0.
\ea
For the string degrees of freedom we obtain the equation for the 
string metric
\be
\lbl{1gamma}
\frac{T_2}{2} 
\sqrt{-\gamma}\left[\partial_iX^M
\partial_j X^Ne^{\left( b+\frac{4}{3}\right)\phi}g_{MN}
-\frac{1}{2}\gamma^{ij}\gamma_{kl}\partial_kX^M\partial_lX^N
e^{\left( b+\frac{4}{3}\right)\phi}g_{MN}\right]=0,
\ee
and for the coordinate fields
\ba
\lbl{1x}
\lefteqn{
T_2\left[\partial_i\left(\sqrt{-\gamma}\gamma{ij}\partial_jX^N
e^{\left( b+\frac{4}{3}\right)\phi}g_{MN}\right)\right.
+\frac{1}{2}\sqrt{-\gamma}\gamma^{ij}\partial_iX^N\partial_jX^P
\partial_M\left(e^{\left( b+\frac{4}{3}\right)\phi}g_{NP}\right)}
\nonumber\\
&&- \left.
3\varepsilon^{ij}\partial_iX^N\partial_jX^PH_{MNP}\right]= 0.
\hspace*{6cm}
\ea

We now follow the standard procedure to find a string soliton. We split up the
coordinates ($M=0,1,\ldots 9$)
\be
\lbl{2}
x^M=(x^\mu,y^m)
\ee
where $\mu=0,1$ and $m=2,\ldots 9$, and make the ansatz
\be
\lbl{3}
ds^2=e^{2A}\eta_{\mu\nu}dx^\mu dx^\nu-e^{2B}\delta_{mn}dy^mdy^n,
\ee
\be
\lbl{4}
B_{\mu\nu}=\gamma\frac{\varepsilon_{\mu\nu}}{\sqrt{g_2}}e^C,
\ee
with $g_2 = -\mbox{det}g_{\mu\nu}$.
All other fields are put equal to zero, and the only coordinate dependance is 
on
$y=\sqrt{\delta_{mn}y^my^n}$. We allow a numerical constant $\gamma$ in the
definition of $B_{\mu\nu}$ to take care of the different normalization compared
to \cite{DKL94}.
The string coordinate $X^M(\xi)$ is split up in the same way as the
coordinates (\ref{2}), and we make the static gauge choice $X^\mu=\xi^\mu$,
and assume $Y^m=\mbox{constant}$, put to zero for simplicity.

The $\gamma_{ij}$ equation now immediately expresses $\gamma_{ij}$
as a function of the metric and the string coordinates, inserting this,
and all the expressions above into the remaining field equations we have
\ba
\lbl{5}
\lefteqn{
\sqrt{-g}e^{4\left( a+\frac{1}{3}\right)\phi} g^{\mu\nu}\left[D^m\partial_m
\left(A+7B+4\left( a+\frac{1}{3}\right)\phi\right) 
\right. 
+\partial A^2-21\partial B^2 }
\nonumber\\
&&
+4\left( a+\frac{1}{3}\right)\partial A
\partial \phi + \left(7a\left(a +\frac{2}{3}\right)+\frac{16}{9}\right)
\partial\phi^2 
+\left.\gamma^2e^{-4A-2\left(a +\frac{4}{3}\right)\phi}
\left(\partial e^{2A+C}\right)^2\right]
\nonumber\\
&&
= \kappa^2T_2e^{2A+\left(b +\frac{4}{3}\right)\phi}g^{\mu\nu}\delta^8(y),
\ea
\ba
\lbl{6}
\lefteqn{
\sqrt{-g}e^{4\left(a +\frac{1}{3}\right)\phi}g^{pm}g^{qn}
\left[
D_m\partial_n(A+3B+2\left(a +\frac{1}{3}\right)\phi)
+\partial_mA\partial_nA +3\partial_mB\partial_nB\right.}
\nonumber\\
&&
-\left(a -\frac{2}{3}\right)\left(a +\frac{4}{3}\right)\partial_m\phi
\partial_n\phi
-\gamma^2e^{-4A-2\left(a +\frac{4}{3}\right)\phi}
\partial_me^{2A+C}\partial_ne^{2A+C}
\nonumber\\
&&
-\frac{g_{mn}}{2}\left[D^p\partial_p(2A+6B + 4\left(a +\frac{1}{3}\right)\phi)
+3\partial A^2-15\partial B^2
\right.
+8\left(a +\frac{1}{3}\right)\partial A \partial\phi 
\nonumber\\
&&
+\left(7a\left(a +\frac{2}{3}\right)+\frac{16}{9}\right)
\partial\phi^2
\left.\left.
-\gamma^2e^{-4A-2\left(a +\frac{4}{3}\right)\phi}
\left(\partial e^{2A+C}\right)^2\right]\right] =0,
\ea
\be
\lbl{7}
\gamma e^{2B}\varepsilon^{\mu\nu}\partial^m\left(e^{-2A+6B+2\left(a 
-\frac{2}{3}\right)\phi}
\partial_me^{2A+C}\right)=-\kappa^2T_2\varepsilon^{\mu\nu}\delta^8(y) ,
\ee
\ba
\lbl{8}
\lefteqn{
\sqrt{-g}e^{4\left( a+\frac{1}{3}\right)\phi}
\left[4\left(a +\frac{1}{3}\right)
D^m\partial_m(2A+7B) + 18a\left(a +\frac{2}{3}\right)D^m\partial_m\phi
\right.  }
\nonumber\\
&&
+12\left(a +\frac{1}{3}\right)(\partial A^2-7\partial B^2)
+36a\left(a +\frac{2}{3}\right)(\partial A\partial\phi 
+\left(a +\frac{1}{3}\right)\partial\phi^2)
\nonumber\\
&&\left.
-2\left(a -\frac{2}{3}\right)\gamma^2e^{-4A-2\left(a +\frac{4}{3}\right)\phi}
\left(\partial e^{2A+C}\right)^2\right]
\nonumber\\&&
=\kappa^2T_2\left(b +\frac{4}{3}\right)e^{2A+\left(a +\frac{4}{3}\right)\phi}
\delta^8(y),
\ea
\be
\lbl{9}
\partial_m\left(e^{2A+\left(b +\frac{4}{3}\right)\phi}\right)
=2\gamma\partial_me^{2A+C}.
\ee                         
Equations (\ref{5}) and (\ref{7}) only give one differential equation each,
proportional to $g^{\mu\nu}$ and $\varepsilon^{\mu\nu}$, respectively, while
(\ref{6}) turns out to have two independent components. The easiest
way to find them is to go to spherical coordinates, $y,\theta_1,\theta_2
\ldots \theta_7$. Then the $yy$-component
will yield one equation, and the $\theta_i\theta_j$-components will yield
one equation proportional to $g_{\theta_i\theta_j}$.

The usual way to solve these equations is to require some unbroken
supersymmetry, and
therefore request that the variations of the spinor fields vanish for
some nonzero supersymmetry parameter. If we do
this we will find one unique solution (modulo some integration constants
without physical significance). The  solution found in this way also
turns out to satisfy the field equations. This is very often the case since
Einsteins equation, the dilaton equation and the Yang-Mills equation can 
be obtained as supersymmetry transformations
of the fermionic equations of motion, which are already identically 
satisfied in our ansatz.
But these equations, as well as the remaining ones, 
still need to be checked explicity since the solution only
keeps half of the supersymmetry.

We now want to show that the equations of motion have a more general solution.
It turns out to be useful to make the field redifinitions
\be
\lbl{10}
X = 2A + 6B + 4\left(a+\frac{1}{3}\right)\phi,
\ee
\be
\lbl{11}
Y = 2A + \left(a-\frac{8}{3}\right)\phi,
\ee
\be
\lbl{12}
Z = 2A + \left(a+\frac{4}{3}\right)\phi.
\ee
The equations of motion can then be written
\ba
\lbl{13}
\lefteqn{
e^{X-2A}\left[\frac{7}{6}(\nabla^2X + \frac{1}{2}X'^2) - \frac{1}{6}
(\nabla^2Y + X'Y' - \frac{1}{2}Y'^2) 
- \frac{1}{2}(\nabla^2Z + X'Z'
\right. }
\nonumber\\
&&\left.
- \frac{1}{2}Z'^2) + \gamma^2 e^{-2Z}\partial\left(e^{2A+C}\right)^2\right]
= -\kappa^2T_2 e^{Z-2A+(b-a)\phi}\delta^8(y),\hspace*{2cm}
\ea
\be
\lbl{14}
e^{X-2B}\left[X'' + \frac{13X'}{y} + X'^2\right] = 0,
\ee
\be
\lbl{15}
e^{X-2B}\left[7\left(X''-\frac{1}{12}X'^2\right) + \frac{13}{12}Y'^2
+ \frac{13}{4}Z'^2 - 13 \gamma^2 e^{-2Z}\left(\partial e^{2A+C}\right)^2
\right] = 0,
\ee
\be
\lbl{16}
\gamma\frac{1}{y^7}\partial\left(y^7e^{X-2Z}\partial e^{2A+C}\right)
= \kappa^2T_2\delta^8(y),
\ee
\ba
\lbl{17}
\lefteqn{
e^X\left[\frac{14}{3}\left(a+\frac{1}{3}\right)(\nabla^2X+\frac{1}{2}X'^2)
-\frac{1}{6}\left(a-\frac{8}{3}\right)(\nabla^2Y + X'Y')
+ \frac{1}{3}\left(a+\frac{1}{3}\right)Y'^2  
\right.}
\nonumber\\
&&\left.
-\frac{1}{2}\left(a+\frac{4}{3}\right)(\nabla^2Z+X'Z') +
\left(a+\frac{1}{3}\right)Z'^2 - 2\gamma^2\left(a-\frac{2}{3}\right)
e^{-2Z}\left(\partial e^{2A+C}\right)^2\right]
\nonumber\\ 
&&
= -\kappa^2T_2\left(b+\frac{4}{3}\right)e^{Z+(b-a)\phi}\delta^8(y),
\ea
\be
\lbl{18}
\partial e^{Z+(b-a)\phi} = 2\gamma\partial e^{2A+C}.
\ee
All derivatives are now with respect to $y$ only, with $\nabla^2 F=
\frac{1}{y^7}\partial(y^7 F')$ and $\partial F = F' = \frac{\partial}
{\partial y} $.

Our strategy is now to solve all equations for $y>0$, and only afterwards 
investigate the singularity structure at $y=0$ to find a consistent
choice for the values of some of the integration constants. We start by
solving the supergravity equations alone.
Equation \ref{14} can be immediately solved
\be
\lbl{19}
e^X = e^{X_0} + \frac{K}{y^{12}}.
\ee
In the following, a subscript zero will always denote a constant which is 
equal to the value of the function in question at infinity, and $K,L \ldots$
are integration constants.
One of the conditions for a supersymmetric solution (see below) is $X'=0$, so 
we immediately see that our nontrivial candidate to a generalization cannot
preserve supersymmetry. 
Now (\ref{16}) can be integrated once yielding
\be
\lbl{21}
e^{X-2Z}\partial e^{2A+C} = \frac{L}{y^7},
\ee
and choosing 
$\gamma = \frac{1}{2}$ (cf (\ref{18}) and Appendix) 
the remaining three equations are equivalent to
\be
\lbl{22}
\nabla^2Y + X'Y' = 0,
\ee
\be
\lbl{23}
\nabla^2Z + X'Z' - e^{2Z-2X}\frac{L^2}{y^{14}} = 0,
\ee
\be
\lbl{24}
7\cdot 12 \frac{Ke^{X_0-2X}}{y^{14}} - \frac{1}{4}\frac{L^2e^{2Z-2X}}{y^{14}}
+\frac{1}{12}Y'^2 + \frac{1}{4}Z'^2 = 0.
\ee
The first of these equations immediately gives
\be
\lbl{25}
e^XY' = \frac{M}{y^7}.
\ee
This can be integated once more, but the result depends on the sign of $K$, so
we will put it off till this sign is known. Instead we first solve (\ref{24}) 
for $Z'$
\be
\lbl{26}
\frac{Z'}
{\left(L^2e^{2Z} - \frac{M^2}{3} - 7 \cdot 48K e^{X_0}\right)^{1/2}}
= \pm \frac{e^{-X}}{y^7}.
\ee
Equation \ref{23} is then identically satisfied provided $Z'\neq 0$, and for 
$Z'=0$ it gives us $L=0$.
We can also directly integrate, (\ref{26}) and afterwards solve for 
$e^{-Z}$ in terms of $y$, for all the values of the integration constants
for which the square root is real. Again we will not do so explicitly, since 
the functional form of the result depends on the signs of the constants.
Combining (\ref{26}) and (\ref{21}) we get
\be
\lbl{260}
\partial e^{2A+C}=\pm \frac{LZ'e^{2Z}}
{\left(L^2e^{2Z} - \frac{M^2}{3} - 7 \cdot 48K e^{X_0}\right)^{1/2}}
\ee
so 
\be
\lbl{261}
e^{2A+C}= \pm
\frac{\left(L^2e^{2Z} - \frac{M^2}{3} - 7 \cdot 48K e^{X_0}\right)^{1/2}}{L}
+\mbox{constant} 
\ee
This completes the solution of the supergravity equations, except for the 
case $L=0$ which will be considered later. Depending on the values of $K$,
$L$ and $M$, these equations have singularities at $y=0$, and we must therefore
add a source term. The string source we have so far neglected gives us
equation \ref{18} which is integrated 
\be
\lbl{20}
e^{2A+C} = e^{Z+(b-a)\phi} +\mbox{constant}.
\ee
The precise value of the constant is not very interesting since it does not
affect $H_{m\mu\nu}$.
We then have
\be
\lbl{28}
e^{Z+\frac{b-a}{4}(Z-Y)} = \pm
\frac{\left(L^2e^{2Z} - \frac{M^2}{3} - 7 \cdot 48K e^{X_0}\right)^{1/2}}{L}
+\mbox{constant}.
\ee 
This can be achieved in two different ways:\footnote{We have not fully
proven that there are no further solutions, but it seems unlikely.}
\begin{itemize}
\item 
The square root proportional to $e^Z$
\ba
(b-a)(Z-Y)= 0
\nonumber\\
M^2+7\cdot144Ke^{X_0} = 0
\nonumber\\
\ea
with the solutions 
\be
\lbl{29}
b=a
\ee
\be
\lbl{30}
K=-\frac{M^2e^{-X_0}}{7\cdot 144}
\ee
or 
\be
Z'=Y'
\ee
\item
The square root constant
\ba
Z=\mbox{constant}
\nonumber\\
Z+\frac{(b-a)(Z-Y}{4})=\mbox{constant}
\nonumber
\ea
\end {itemize}
Both $Z'=Y'$ and $Z'=0$ have as a consequence $L=0$. This case will be treated 
in Section~4.

\sect{The heterotic string solution} 

In this section, we will study the solution for $L\neq 0$ given in Section~2, 
and show that this is the one corresponding to the elementary string 
solution of \cite{DGHR90}. That it is indeed a heterotic string was shown in
\cite{Dab95,H195}, where the zero modes are fully analyzed.

We insert (\ref{29}) and (\ref{30}) in our equations and 
find
\be
\lbl{31}
Y' = Me^{-X_0}\frac{y^5}{y^{12}-\frac{M^2e^{-2X_0}}{7\cdot 144}}
\ee
and 
\be
\lbl{33}
\partial e^{-Z} = 
-L e^{-X_0}\frac{y^5}{y^{12}-\frac{M^2e^{-2X_0}}{7\cdot 144}}.
\ee
This can now be integrated to
\be
\lbl{32}
Y =  \left\{\begin{array}{ll}Y_0 +
\frac{\sqrt{7}M}{|M|}\log \left|\frac{y^6-\frac{|M|e^{-X_0}}{\sqrt{7}\cdot 12}}
{y^6 +\frac{|M|e^{-X_0}}{\sqrt{7}\cdot 12}}\right| & M \neq 0
\\
Y_0 & M=0
\end{array}
\right.
\ee
and
\be
\lbl{34}
e^{-Z} = \left\{\begin{array}{ll} e^{-Z_0} -
\frac{\sqrt{7}L}{|M|}\log \left|\frac{y^6-\frac{|M|e^{-X_0}}{\sqrt{7}\cdot 12}}
{y^6 + \frac{|M|e^{-X_0}}{\sqrt{7}\cdot 12}}\right| & M \neq 0
\\
e^{-Z_0} + \frac{Le^{-X_0}}{6y^6} & M=0
\end{array}
\right. .
\ee
A study of the singularities at $y=0$ yields
\be
\lbl{35}
e^XY' = M\Omega_7 f'
\ee
\be
\lbl{36}
e^X\partial e^{-Z} = -L\Omega_7 f'
\ee
where $\nabla^2 f = \delta^8(y)$, and $\Omega_7$ is the volume of the 
seven-sphere. The nonvanishing parts of the equations of motion at $y=0$
are then
\be
\lbl{37}
e^{-2A}\left[-\frac{1}{6}M - \frac{1}{2}Le^Z\right]\Omega_7\delta^8(y)
= -\kappa^2T_2 e^{Z-2A}\delta^8(y),
\ee
\be
\lbl{38}
L\Omega_7\delta^8(y) = 2\kappa^2 T_2 \delta^8(y),
\ee
\be
\lbl{39}
\left[-\frac{1}{6}\left(a-\frac{8}{3}\right)M - \frac{1}{2}\left(a+\frac{4}{3}
\right)Le^Z\right]\Omega_7\delta^8(y) = - \kappa^2T_2\left(a+\frac{4}{3}\right)
e^Z \delta^8(y),
\ee
So the constants must take the values\footnote{$e^{Z(0)}\neq 0$ for $M\neq 0$}
\be
\lbl{40}
L = \frac{2\kappa^2T_2}{\Omega_7},
\ee
\be
\lbl{41}
M = 0.
\ee
Choosing $M=0$ we 
get 
\be
\lbl{491}
e^{Z(0)}=0,
\ee
\be
\lbl{492}
e^{Z(0)-2A(0)}=e^{\frac{1}{4}\left(a+\frac{4}{3}\right)(Z(0)-Y(0))}.
\ee
Here $Y(0)=Y_0$, so the factor multiplying the $\delta$-function in
(\ref{37}) is divergent for $a+\frac{4}{3}<0$, and zero for $a+\frac{4}{3}>0$.
This is just a curiosity here, but 
will be important for the solitonic string solution.

The remaining integration constants are the values of the fields at infinity,
$X_0$, $Y_0$ and $Z_0$, which can be rewritten in terms of $A_0$, $B_0$ and 
$\phi_0$. The first two of these have no physical meaning. They can be removed
by constant rescaling of the coordinates, so we are left with only $\phi_0$,
which is the vacuum expectation value of the dilaton field.
This is exactly the standard elementary string solution of \cite{DGHR90,DKL94},
written 
for genaral $a$, that is, with all possible rescalings of the metric explicitly
given. In order to compare directly One should let $a=-\frac{1}{3}$, and solve 
$X$ and $Y$ for $A$ and $B$ in terms of $\phi$. 
We have a preserved N\oe ther charge 
\be
\lbl{42}
e = \frac{6}{\sqrt{2}\kappa}\int e^{2\left(a+\frac{2}{3}\right)\phi} *H 
= \sqrt{2}\kappa T_2.
\ee
The factor 6 is put in to have the same definition as \cite{DKL94} using our 
normalization of $H$.
The mass per unit string length is (assuming $g_{MN}\rightarrow
e^{2A_0}\eta_{MN}$ for $y\rightarrow \infty$, and keeping $A_0\neq 0$ as an 
extra check on the calculation.)
\ba
\lbl{43}
{\cal M}_2 = -\frac{1}{2\kappa^2} e^{6A_0}\int d^8 y 
e^{4\left(a+\frac{1}{3}\right)\phi}\nabla^2 \left(e^{2(A-A_0)} + 7 e^{2(B-A_0)}
\right)
\nonumber\\
= 2\left(a+\frac{5}{6}\right)T_2 e^{\left(a+\frac{4}{3}\right)\phi_0}
\ea
Here we can see explicitly that ${\cal M}_2$  scales with $g_{MN}$ as 
predicted in \cite{H295}. The fact that
${\cal M}_2$ is negative in, for instance, the heterotic string metric,
$a=-\frac{4}{3}$, should be no cause for worry as long as it can be interpreted
as a (positive) energy density in the Einstein metric, $a=-\frac{1}{3}$.

The conditions that the solution preserve half the supersymmetry can 
be obtained just like in the papers quoted above. In our notation they are 
\be
\lbl{44}
3\left(A' + \frac{a}{2}\phi'\right) = e^{-Z}\partial e^{2A+C}
\ee
\be
\lbl{45}
6\left(B' + \frac{a}{2}\phi'\right) = - e^{-Z}\partial e^{2A+C}
\ee
\be
\lbl{46}
4\phi' = e^{-Z}\partial e^{2A+C}
\ee
which can be rewritten as
\be
\lbl{47}
X' = 0
\ee
\be
\lbl{48}
Y'= 0
\ee
\be
\lbl{49}
\partial e^{Z} = \partial e^{2A+C}.
\ee
These equations are all satisfied here as we know they should be.

The solution we have found can also be interpreted as a solitonic
string solution of the dual version of ten-dimensional supergravity, 
nowadays usually
interpreted as the five-brane theory
\cite{DL291}. We replace $H$ by $\tilde{H}=
e^{2\left(a-\frac{2}{3}\right)\phi}*H$.
All supergravity equations remain unchanged, but the $H$-equation 
of motion is now interpreted as the Bianchi identity for $\tilde{H}$, and the 
Bianchi identity for $H$ becomes the equation of motion for the new
field strength. This equation has no singularity at $y=0$, and we have
no source term. The constant $L$ is then no longer fixed by
the equations at $y=0$\footnote{It still has to fulfill a Dirac quantization
condition of the form $e_6\mu_2=2\pi n$, see \cite{DKL94}.}.
It gives us the conserved charge
\be
\lbl{490}
\mu_2 = \frac{L\Omega_7}{\sqrt{2}\kappa}
\ee
which is now interpreted as a magnetic charge. However, the solution still
has to satisfy (\ref{37}) and (\ref{39}) with zero on the rhs.
For $M\neq 0$ the exponentials have finite values at $y=0$, and we can only get 
the solution $L=M=0$ in contradiction with the assumption. 
We then have to choose $M=0$. Equations \ref{491} and \ref{492} then tell
us that 
we can have a non-zero $L$, and hence a non-trivial 
solution only for 
\be
\lbl{493}
a+\frac{4}{3} > 0.
\ee
This is the standard solitonic solution again satisfying the conditions for 
unbroken supersymmetry (\ref{47})-(\ref{49}).
In principle there is no reason to require (\ref{18}) in this case. If we 
then study the singularity structure at $y=0$ we find that we must still have 
$M=0$, but that the other condition is generalized to $\left(L^2e^{2Z(0)}-
\frac{M^2}{3}-7\cdot 48 Ke^{X_0}\right)^{1/2}=0$. If we solve for $e^{-Z}$,
and require the solution to exist and be well-behaved on the full interval
$0\leq y <\infty $
the condition can not be satisfied for $K\neq 0$.

Notice that while the solitonic string solution exists in the type~I metric,
in the Einstein metric, and in the five-brane metric, there is no solution 
with $L\neq 0$  in the
heterotic string metric, $ a=-\frac{4}{3}$. This is probably an indication
that the heterotic string is a fundamental state of ``heterotic" supergravity,
and not a solitonic one. 

The supersymmetric solution to the supergravity equations 
has an extension to $\alpha'\neq 0$.
If we use the framework
of anomaly free supergravity, given explicitly in \cite{P91}, 
where the Green-Schwarz anomaly cancellation
term is supersymmetrized in a consistent fashion, we find an exact solution
to the equations of motion reducing to the one above for $\alpha' = 0$
\cite{T96}. 
Assuming that there exists an action (probably containing infinitely many
terms) for these equations of motion, we can ask ourselves what source
terms we must have at $y=0$, just like we did for the elementary string 
solution above. We will find that also
the string action has to be modified with higher order terms, which might 
possibly be removable by an $\alpha'$-dependent field redifinition. In 
\cite{PW95} it is argued that this elementary heterotic string 
is really a D-string in strongly coupled type~I theory, since their world 
sheet structures are the same. This will then give us the exact conformal
field theory construction corresponding to the full solution.

\sect{The type~I string solution}
The second possible way of finding a solution requires
$L=0$. In this case the lhs of equation \ref{16} has no singularity at
$y=0$ and hence we cannot have a source at the rhs. We then have to redo
the analysis putting $\partial e^{2A+C}=0$ in (\ref{13}), (\ref{15}), 
(\ref{17}) and
(\ref{18}), and removing the rhs of (\ref{16}). In this case we will not
find an explicit solution for $b$, but if we put $b=a-2$  by hand, we find 
that the solution can be interpreted as an unstable type~I configuration,
which should be there if there is indeed a strong-weak coupling duality between
ten dimensional heterotic and type~I strings.

We first solve equations \ref{13}-\ref{17} for $y>0$. Equation \ref{14}
gives the same solution for $e^X$, and the remaining equations are now
\be
\lbl{50}
\nabla^2Y + X'Y' = 0,
\ee
\be
\lbl{51}
\nabla^2Z + X'Z' = 0,
\ee
\be
\lbl{52}
\partial e^{2A+C} = 0,
\ee
\be
\lbl{53}
7\cdot 12\frac{Ke^{X_0-2X}}{y^{14}} + \frac{Y'^2}{12} +\frac{Z'^2}{4} =0,
\ee
with the solutions (for $K$ different from zero)
\be
\lbl{54}
Y = Y_0 + \frac{Me^{-X_0/2}}{12\sqrt{-K}}\log \left|
\frac{y^6-\sqrt{-K}e^{-X_0/2}}{y^6+\sqrt{-K}e^{-X_0/2}}\right|  ,
\ee
\be
\lbl{55}
Z = Z_0 + \frac{Ne^{-X_0/2}}{12\sqrt{-K}}\log \left|
\frac{y^6-\sqrt{-K}e^{-X_0/2}}{y^6+\sqrt{-K}e^{-X_0/2}}\right| ,
\ee
\be
\lbl{56}
K = - \frac{M^2 + 3N^2}{7\cdot 144}e^{-X_0}.
\ee
At $y=0$ we now have 
\be
\lbl{57}
\left[-\frac{1}{6}M -\frac{1}{2}N\right] \Omega_7\delta^8(y)
= - \kappa^2T_2e^{Z-2A+(b-a)\phi}\delta^8(y),
\ee
\be 
\lbl{58}
\left[-\frac{1}{6}\left(a-\frac{8}{3}\right)M - \frac{1}{2}\left(a+\frac{4}{3}
\right)N\right]\Omega_7\delta^8(y) = -\kappa^2T_2\left(b+\frac{4}{3}\right)
e^{Z+(b-a)\phi}\delta^8(y).
\ee
These equations have the solutions
\be
\lbl{59}
M = -\frac{3}{2}\frac{\kappa^2T_2}{\Omega_7}(b-a)e^{(Z+(b-a)\phi)(0)},
\ee
\be
\lbl{60}
N =  \frac{1}{2}\frac{\kappa^2T_2}{\Omega_7}(b-a+4)e^{(Z+(b-a)\phi)(0)}.
\ee
This is in contradiction with (\ref{18}), which, in this case,
requires that $(b-a+4)^2+3(b-a)^2=0$. However, equation \ref{18} is the
equation of motion for the string source. 
It can be interpreted as a
no-force condition, stating that the graviton contribution to the force
between two parallell source strings is cancelled by the ``axion'' 
contribution. In the present case we have no axion contribution (to the 
lowest order in $\alpha'$ at least). Furthermore, this solution does not
preserve supersymmetry, nor does it have a conserved N\oe ther charge,
so there is no reason to expect it to be stable. It is hence just the 
configuration around a source term corresponding to an 
infinitely long string put in by hand. 
There is no reason 
to expect that such an  unstable test source should satisfy dynamical
equations of motion, or that there should be no force between two parallell 
unstable strings, so we just drop
equation \ref{18}.

The reason we are still interested in this solution is that we 
are looking
for just such a thing as the missing type~I ``soliton", \cite{H295}. 
The mass per unit string length for our solution is
\be
\lbl{61}
{\cal M}_2 = \frac{T_2}{2}\left[3\left(a+\frac{1}{6}\right)(b-a) +
\left(a+\frac{5}{6}\right)(b-a+4)\right]e^{\left(b+\frac{4}{3}\right)\phi_0}
\ee
If we choose $b=a-2$, corresponding to a type~I string source we obtain 
\be
\lbl{62}
{\cal M}_2 = -2\left(a-\frac{1}{2}\right)e^{\left(a-\frac{2}{3}\right)\phi_0}.
\ee
This is indeed the correct scaling behaviour for a type~I soliton, 
since $a=\frac{2}{3}$ in the type~I metric. The fact that there is no
$B_{MN}$ term in the source is also consistent with the type~I interpretation
since $B_{MN}$ is here a Ramond-Ramond state corresponding to 
a sigma model term $\gamma^{MNP}H_{MNP}$ sandwiched between two spin fields, 
see \cite{CLNY88}, and such a term should vanish in
our ansatz.

Can we also here reinterpret the solution as a ``solitonic" type~I string, 
that is a solution where we have exchanged the role of the equation of
motion and the Bianchi identity for $H$, 
see Section~2, and with no sources at $y=0$?
We find
\be
\lbl{620}
\tilde{H}_{\theta_1\theta_2\theta_3\theta_4\theta_5\theta_6\theta_7}
\sim \sqrt{h} y^7 e^{X-2Z}\partial e^{2A+C}
\ee
with $h_{\theta_i\theta_j}$ the metric on $S^7$ and 
$h=\det h_{\theta_i\theta_j}$. The solution of the Bianchi identity now
requires that $\tilde{H}$ be independent of $y$, and we are back at (\ref{21}).
With $L=0$ we then have exactly the same equations as we already studied above,
with the only difference that (\ref{58}) and (\ref{59}) have zero on the
rhs. The only solution is then $M=N=0$, so, as should have been expected,
there is no interpretation as a 
solitonic string in this case.

As we already mentioned it has also been shown that the zero modes
for the elementary string case 
indeed represent the degrees of freedom of a heterotic string 
\cite{Dab95,H195}.
That
analysis certainly does not work for the case of the non-supersymmetric 
solutions of the supergravity equations, since we have twice the wanted number 
of fermionic zero-modes. 

Since, by our analysis, nonsupersymmetric, unstable solutions also exist 
around non-dynamical heterotic string sources put in by hand, it feels very
tempting to believe that our solution to the supergravity equations 
should also contain an exact type~I string solution (Even without invoking 
the duality arguments.) What we need is a
non-vanishing field strength, providing a conserved charge, and a source term
able to counteract the gravitational forces between two string sources.
As I can see it there are two possiblilities. The immediate one is that we
have to employ a standard $B_{MN}$ term in the string action to higher orders 
in $\alpha'$, so that the good solution, which can then probably only be
found if we solve the equations to all orders in $\alpha'$, vanishes, or 
becomes singular, for $\alpha'\rightarrow 0$. 
It might be difficult to find such a solution, which preserves some of the
supersymmetry, however, since we must satisfy (the generalization of) 
(\ref{49}).
The second, perhaps more amusing 
possibility is that the Yang-Mills field will provide us with what we need.
There are obvious difficulties with this possiblity too. Solitons 
with Yang-Mills charge are 0-branes, not strings, and they also turn out to
be non-supersymmetric \cite{zerob}. If the 0-branes could somehow be 
interpreted as
type~I strings of zero length, one part of the problem would be solved, but we
would still need to find the unbroken supersymmetry. (And show that the 
zero-modes of the solution provides us with the degrees of freedom for a type~I
string).

Our general solution also permits $b$ to take any value, as long as the 
string is not dynamical. It is hard to think of any further 
physical interpretation
of any of these solutions. The interpretation is probably just that 
one can find an exact supergravity configuration around many unusual sources
put in by hand, while if you want an exact solution also to 
the string equations our equations forced us to put $b=a$.
We then expect that a stable type~I solution should force us to choose $b=a-2$.

\sect{The general case}
Our general (non-supersymmetric?) solution actually exists for 
generic dimensions. We will briefly give this solution here without any
attempts at interpretation.

We start with the action of a rank $d$ antisymmetric tensor potential
interacting with gravity and the dilaton in $D$ space-time dimensions,
and couple it to a $p=d$ $p-1$-brane. The procedure is exactly analogous to the 
$D=10$, $d=2$ case we already did in great detail. Here we choose the Einstein
metric so we use the same action (and equations of motion) as in
\cite{DKL94} (except for the minor redefinitions of the 
fields, see Appendix). Splitting up the coordinates in $D=d + (\tilde{d}+2)$, 
and making the usual ansatz the equations corresponding to 
(\ref{13})-(\ref{18})
are now
\ba
\lbl{101}
e^{X-2A}\left[\frac{\tilde{d}+1}{\tilde{d}}(\nabla^2X + \frac{1}{2}X'^2) 
- \frac{a^2(d+\tilde{d})}{Nd\tilde{d}}
(\nabla^2Y + X'Y' - \frac{1}{2}Y'^2) 
- \frac{2}{N}(\nabla^2Z 
\right. 
\nonumber\\
\left. + X'Z'
- \frac{1}{2}Z'^2) + \frac{1}{4} e^{-2Z}\partial\left(e^{2A+C}\right)^2\right]
= -\kappa^2T_d e^{Z-2A+(b-a)\phi}\delta^{\tilde{d}+2}(y),
\ea
\be
\lbl{102}
e^{X-2B}\left[X'' + \frac{(2\tilde{d}+1)X'}{y} + X'^2\right] = 0,
\ee
\ba
\lbl{103}
\lefteqn{
e^{X-2B}\left[(\tilde{d}+1)\left(X''-\frac{1}{2\tilde{d}}X'^2\right) 
\right.}
\nonumber
\\
&&+ \left.
(2\tilde{d}+1)\left[\frac{a^2(d+\tilde{d})}{2Nd\tilde{d}}Y'^2
+ \frac{1}{N}Z'^2 -  \frac{1}{4} e^{-2Z}\left(\partial e^{2A+C}\right)^2
\right]\right] = 0,
\ea
\be
\lbl{104}
\frac{1}{y^{\tilde{d}+1}}\partial\left(y^{\tilde{d}+1}e^{X-2Z}
\partial e^{2A+C}\right)
= \kappa^2T_d\delta^{\tilde{d}+2}(y),
\ee
\ba
\lbl{105}
e^X\left[ -\frac{2a}{N}(\nabla^2Y + X'Y')
+\frac{2a}{N}(\nabla^2Z+X'Z') 
- \frac{a}{2}e^{-2Z}\left(\partial e^{2A+C}\right)^2\right]
\nonumber\\
= -\kappa^2T_de^{Z+(b-a)\phi}\delta^{\tilde{d}+2}(y),
\ea                                                  
\be
\lbl{106}
\partial e^{Z+(b-a)\phi} = \partial e^{2A+C}.
\ee

Here 
\be
\lbl{107}
X = dA+\tilde{d}B,
\ee
\be
\lbl{108}
Y = dA-\frac{2d\tilde{d}}{a(d+\tilde{d})}\phi,
\ee
\be
\lbl{109}
Z = dA + a\phi
\ee
and $N=a^2+\frac{2d\tilde{d}}{d+\tilde{d}}$. The constants $a$ and $b$
here are really regarded as unknown, to be solved for. They have nothing to
do with rescaling the metric since the equations are all written in the 
Einstein metric, but instead determine how the $d$-form potential couples to
the dilaton.

We will now do the case where the source term does satisfy dynamical equations,
to show it is exactly analogous to Section~2. Again we first study 
the supergravity part for $y\neq 0$ so
we can put the rhs to zero in (\ref{101}), (\ref{104}) and (\ref{105}).
First we solve equation \ref{102}
obtaining
\be 
\lbl{110}
e^X = e^{X_0} + \frac{K}{y^{2\tilde{d}}},
\ee
We integrate (\ref{104}) once to get
\be
\lbl{112}
e^{X-2Z}\partial e^{2A+C} = \frac{L}{y^{\tilde{d}+1}}.
\ee
In exact analogy to the string case the remaining three equations then turn out
to be equivalent to
\be
\lbl{113}
\nabla^2(aY) + X'(aY') = 0,
\ee
\be
\lbl{114}
\nabla^2Z + X'Z' - \frac{N}{4}e^{2Z-2X}\frac{L^2}{y^{2(\tilde{d}+1)}} = 0,
\ee
\be
\lbl{115}
2\tilde{d}(\tilde{d}+1)\frac{Ke^{X_0-2X}}{y^{2(\tilde{d}+1)}} 
-\frac{1}{4}\frac{L^2e^{2Z-2X}}{y^{2(\tilde{d}+1)}}
+ \frac{(d+\tilde{d})}{2d\tilde{d}N}(aY')^2 + \frac{1}{N}Z'^2 = 0,
\ee
where we must use $aY$ instead of $Y$ to have a field which is regular
also for $a=0$. The solution to equation \ref{113}, 
\be
\lbl{116}
e^X(aY') = \frac{M}{y^{\tilde{d}+1}}, 
\ee
can be inserted into (\ref{115}) which is then solved for $Z'$.
Just like in the string case this $Z'$ identically satisfies (\ref{114}).
We find 
\be 
\lbl{118}
\frac{Z'}{\left(e^{2Z}\frac{NL^2}{4}-\frac{(d+\tilde{d})M^2}{2d\tilde{d}}
-2N\tilde{d}(\tilde{d}+1)Ke^{X_0}\right)^{1/2}}
 = 
\pm \frac{e^{-X}}{y^{\tilde{d}+1}}
\ee
Both (\ref{116}) and (\ref{118}) can be integrated, and $Y$ and 
$e^{-Z}$ expressed in terms of y. Combining (\ref{118}) with (\ref{112}) we 
find
\be 
\lbl{1111}
e^{2A+C} = \pm \frac{4}{NL}
\left(e^{2Z}\frac{NL^2}{4}-\frac{(d+\tilde{d})M^2}{2d\tilde{d}}
-2N\tilde{d}(\tilde{d}+1)Ke^{X_0}\right)^{1/2}
+ \mbox{constant}.
\ee

We will now only study the solution to this equation consistent with the
full rhs. 
Integrating (\ref{106}) gives
\be
\lbl{111}
e^{2A+C} = e^{Z+(b-a)\phi} +\mbox{constant}.
\ee
We must then choose
\be
\lbl{119}
b=a,
\ee
\be
\lbl{120}
K = -\frac{(d+\tilde{d})M^2e^{-X_0}}{4Nd\tilde{d}^2(\tilde{d}+1)}
\ee
\be
\lbl{117}
N=4
\ee
Then we have
\be
\lbl{121}
e^{-Z} = \left\{\begin{array}{ll} e^{Z_0} -
\sqrt{\frac{d(\tilde{d}+1)}{d+\tilde{d}}}\frac{2L}{|M|}
\log \left|\frac{y^{\tilde{d}}- \sqrt{\frac{d+\tilde{d}}{d(\tilde{d}+1)}}
\frac{|M|e^{-X_0}}{4\tilde{d}}}
{y^{\tilde{d}}+ \sqrt{\frac{d+\tilde{d}}{d(\tilde{d}+1)}}
\frac{|M|e^{-X_0}}{4\tilde{d}}}
\right| & M \neq 0
\\
e^{-Z_0} + \frac{Le^{-X_0}}{\tilde{d}y^{\tilde{d}}} & M=0
\end{array}
\right. .
\ee
If we now continue and analyze the equations at $y=0$ we will again find
that we have to put in a $p-1$-brane source term at the origin getting
\be
\lbl{122}
L=\frac{2\kappa^2T_d}{\Omega_{\tilde{d}+1}},
\ee
\be
\lbl{123}
K = M =0,
\ee
which is the elementary solution found in \cite{DKL94}. Alternatively we 
can again examine the dual solution and again find the result of \cite{DKL94}.

Also in the general case we have to study $L=0$ separately.
The analysis of this case closely follows Section~4, only yielding 
unstable solutions. We will not show these solutions explicitly here since 
the only interesting interpretation we know of is the type~I string solution
already considered.

\sect{Conclusions}

We have have found a general solution to the supergravity equations 
with the
standard ansatz for all fields. A specific choice of integration constants 
gives us the known supersymmetric elementary and solitonic $p-1$-branes.
These are the only solutions with a dynamical $p-1$-brane source, and with no 
source term. We also show that there is a string solution with a fixed
source giving us a probably unstable  type~I configuration.

Making other choices for the integration constants, will give us other  
solutions, were the singularities at $y=0$ can be accounted for
with a source term with an arbitrary relative constant between
the $g_{MN}$-term and the $B_{MN}$-term, as well as an arbitrary $b$.
There is still a possibility that some
other such choice will give us a solution with
a reasonable physical interpretation.

It would now be very interesting to see if we could somehow relax some
of the assumptions here to find a stable, or at least less unstable type~I
solution. What we really would like is of course a solution where the 
zero-modes
match the degrees of freedom of the type~I string. Only then can we feel really
convinced that what we have found is indeed the type~I state in heterotic 
string
theory. We discussed some tentative ideas about how to achieve this at
the end of Section 4, but,
in view of the discussion in \cite{H295}, 
it might, unfortunately, turn out to be impossible to find such 
a state at all. 
\\
\\
{\large{\bf{Acknowledgement}}}
I wish to thank L\'{a}rus Thorlacius and Igor Pesando for helpful discussions.
\newpage
\appendix
\setcounter{equation}{0}
\renewcommand{\theequation}{A.\arabic{equation}}
\noindent
{\Large{\bf{Appendix}}}\\
\\
We use the conventions of \cite{P91} for the fields, and of \cite{DKL94}
for the coupling constants with, for instance
\ba
\lbl{a1}
\{\gamma_M,\gamma_N\} = 2 g_{MN} ,
\nonumber \\
g_{MN}= (+,-,-\ldots)   ,
\ea
and for the string case, Section~2--Section~4,
our action and equations of motion are the ones from these papers 
(with $\gamma_1 \sim \alpha' = 0$, and $R=R(\omega)$, and with $g_{MN}$ 
rescaled to $e^{a\phi}g_{MN}$. 
These conventions
are the ones which came
out naturally in the superspace derivation of \cite{P91}, and will later make
it easier to find solutions consistent to all orders in $\alpha'$.
In order to compare to, for instance, \cite{DKL94} we put $a=\frac{1}{3}$
to obtain the Einstein metric, and find
\be
\lbl{a2}
R=\frac{1}{2}\tilde{R}
\ee
\be
\lbl{a3}
g_{MN}= - \tilde{g}_{MN}
\ee
\be
\lbl{a4}
B_{MN}=\frac{1}{2}\tilde{B}_{MN}
\ee
\be
\lbl{a5}
H_{MNP}=\frac{1}{6}\tilde{H}_{MNP}
\ee
\be
\lbl{a6}
\phi =\frac{1}{2}\tilde{\phi}
\ee
\be
\lbl{a61}
2A+C = \tilde{C}
\ee
where the fields with a tilde are the ones of \cite{DKL94}.
In the generic case we use the Einstien metric only, so here we are much 
closer to this report, only using the generalizations of 
(\ref{a2})-(\ref{a61}),
as our fields.

Finally, we give some explicit expressions using the ansatz 
(\ref{2})-(\ref{4}).
\be
\lbl{a7}
R_{\mu\nu} = \frac{1}{2}g_{\mu\nu}[D^m\partial_mA 
+ 2\partial^mA\partial_mA]
\ee
\be
\lbl{a8}
R_{mn} = D_m\partial_nA + \partial_mA\partial_nA + 3D_m\partial_nB
+\frac{1}{2}g_{mn}D^p\partial_pB + 3 \partial_mB\partial_nB 
-3g_{mn}\partial^p B\partial_p B
\ee
where
\be
\lbl{a9}
D_m\partial_n F = \partial_m\partial_n F - \Gamma^p_{mn}\partial_pF
= \partial_m\partial_nF - \partial_mB\partial_n F - \partial_n B\partial_mF
+g_{mn}\partial_pB\partial^pF
\ee
\be
\lbl{a10}
D_\mu\partial_\nu F = -\Gamma^m_{\mu\nu}\partial_m F = g_{\mu\nu}
\partial^mA\partial_m F
\ee
The generalizations to generic dimension can easily be found from the 
expressions given in \cite{DKL94}.

\end{document}